\documentclass[12pt]{iopart}
\usepackage{graphicx}
\usepackage{iopams} 

\begin{document}
\newcommand{\up}{\uparrow}
\newcommand{\dn}{\downarrow}
\newcommand{\be}{\begin{equation}}
\newcommand{\ee}{\end{equation}}
\newcommand{\bea}{\begin{eqnarray}}
\newcommand{\eea}{\end{eqnarray}}

\title[Polarization in a three-dimensional Fermi gas with Rabi coupling]
{Polarization in a three-dimensional Fermi gas with Rabi coupling}

\bigskip
\author{V Penna$^1$ and L. Salasnich$^{2,3}$}

\address{
$^1$ Politecnico di Torino, 
Dipartimento di Scienza Applicata e Tecnologia, and CNISM, Corso Duca degli Abruzzi 24, 
I-10129 Torino, Italy}

\address{
$^2$ Dipartimento di Fisica e Astronomia ''Galileo Galilei" and CNISM, Universit\'a di
Padova, Via Marzolo 8, I-35131 Padova, Italy}

\address{
$^3$ Istituto Nazionale di Ottica (INO) del Consiglio Nazionale delle Ricerche (CNR),
Via Nello Carrara 1, I-50019 Sesto Fiorentino, Italy}

\begin{abstract}
We investigate the polarization of a two-component three-dimensional fermionic gas  
made of repulsive alkali-metal atoms. The two pseudo-spin components 
correspond to two hyperfine states which are Rabi coupled. 
The presence of Rabi coupling implies that only the total number 
of atoms is conserved and a quantum phase transition between 
states dominated by spin-polarization along different axses is possible. 
By using a variational Hartree-Fock 
scheme we calculate analytically the ground-state energy of the system and 
determine analytically and numerically 
the conditions under which there is this quantum phase transition. 
This scheme includes the well-known criterion for the Stoner instability.
The obtained phase diagram clearly shows that the polarized phase 
crucially depends on the interplay among the Rabi coupling energy, 
the interaction energy per particle, and the kinetic energy per particle. 
\end{abstract}



\section{Introduction}

Few years ago, seminal experiments realized artificial spin-orbit 
and Rabi couplings in bosonic \cite{so-bose,so-bose2} and fermionic 
\cite{so-fermi1,so-fermi2} atomic gases. In these experiments, 
laser beams were used to couple two internal hyperfine states of the atom by 
a stimulated two-photon Raman transition \cite{so-bose,so-bose2,so-fermi1,so-fermi2}. 
Driven by these experiments, many papers have analyzed 
spin-orbit effects in Bose-Einstein condensates 
\cite{stringa1,stringa2,burrello,so-solitons1,so-solitons2,so-solitons3,so-bright1,so-bright2,so-bright3} 
and also in the BCS-BEC crossover of superfluid fermions 
\cite{shenoy1,shenoy2,gong,hu,iskin,dms1,dms2,sa,chen,zhou2,yangwan}.

Very recently, ferromagnetic instability has been reported 
in a two-component three-dimensional (3D) Fermi gas made of 
ultracold $^6$Li atoms \cite{roati}. 
The repulsive interaction between fermionic atoms induces 
the well-known Stoner instability \cite{stoner} above a critical strength. 
In the absence of the Rabi coupling, the system remains balanced and   
this instability produces { balanced ferromagnetism with} 
phase separation rather than spin flip \cite{itinero0,itinero1,itinero2,itinero3,itinero4,
itinero5}. In previous papers \cite{sala-previous,sala-njp2017} 
we have investigated the role of the Rabi coupling in a 2D Fermi gas  
to the formation of spin-flip, i.e. polarization \cite{itinero1}. 
In this 2D theoretical study the density of states is quite simple 
and the obtained phase diagram is fully analytical \cite{sala-njp2017}. 

In this paper we study the emergence of polarization along the $z$ axis 
in a Rabi-coupled 3D Fermi gas of repulsive alkali-metal atoms. Unlike the 2D case, 
the 3D density of states is quite complex and requires an analytical study 
of the equation determining the ground state supported by numerical calculations. 
The fermionic atoms are characterized by 
two hyperfine internal states, which can be modelled as two spin components. 
We neglect the role of molecules which could be relevant in the presence of strong
spin-component interaction.
We analyze the ground-state properties of the quantum gas by using 
the variational Hartree-Fock method, where the population imbalance 
is a variational parameter. We determine analytically and, 
in some regions of the phase diagram, numerically the conditions under 
which there is a quantum transition from a phase dominated by spin-polarization
along the $x$ axis to a spin-polarized one along the $z$ axis. 
We find that this quantum phase transition, 
which corresponds to a spontaneous symmetry breaking of the 
fermion polarization (population imbalance) between 
two degenerate values, appears at a critical interaction strength 
which depends on both the Rabi energy and the kinetic energy. 

After rapidly reviewing, in section \ref{sez2}, the application of the 
mean-field Hartree-Fock method to the model Hamiltonian, in section \ref{sez3}, 
we present the central result of our study,
the phase diagram of the 3D model. We show that two distinct regimes characterize the
system: in the first regime the 3D density of states features a simple form which allows one
to analytically derive the critical line separating the balanced from polarized phase.
Conversely, in the second regime, the 3D density of states features a strong nonlinear
dependence on the significant physical parameters and the derivation of the critical line 
requires a detailed analysis of the ground-state equation and the numerical solution thereof. 
This regime is discussed in section \ref{sez4}, while section \ref{sez5} is devoted to 
the polarization properties of the system.


\section{Field-theory Hamiltonian with Rabi coupling}
\label{sez2}

A 3D fermionic gas 
including contact interaction and Rabi coupling 
is described by the model Hamiltonian
%
\be
{\hat H} = \int d^3 {\bf r} \left [ 
- \sum_{\sigma=\uparrow,\downarrow} 
\frac{\hbar^2}{2m}
{\hat \psi}_{\sigma}^+ \nabla^2 {\hat \psi}_{\sigma} 
+ g \ {\hat n}_{\uparrow}^+  {\hat n}_{\downarrow} 
+
\frac{\hbar \Omega}{2} \left( 
{\hat \psi}_{\uparrow}^+ {\hat \psi}_{\downarrow} 
+ {\hat \psi}_{\downarrow}^+ {\hat \psi}_{\uparrow} 
\right) 
\right ] \; , 
\label{H2}
\ee
where ${\hat n}_{\sigma} = {\hat \psi}_{\sigma}^+ {\hat \psi}_{\sigma}$ 
is the local number density operator for atoms with spin 
$\sigma=  {\uparrow}, {\downarrow} $, $g$ and $\Omega$ are the contact 
interaction and the Rabi coupling, respectively, 
and ${\hat \psi}_{\sigma}({\bf r})$ is the field operator which 
destroys a fermion of spin $\sigma$ at position ${\bf r}$.
A good quantum number associated to $\hat H$ is
the total fermion number  
\be 
{\hat N} = {\hat N}_{\up}+ {\hat N}_{\dn}= \int \! d^3{\bf r} \left( 
{\hat \psi}_{\uparrow}^+({\bf r}) {\hat \psi}_{\uparrow}({\bf r}) + 
{\hat \psi}_{\downarrow}^+({\bf r}) {\hat \psi}_{\downarrow}({\bf r})
\right) 
\ee
since $[{\hat N},{\hat H}]=0$. Conversely,  ${\hat N}_{\up}$ 
and ${\hat N}_{\dn}$, the spin-up and spin-down relative numbers, do not represent conserved quantities 
because the Rabi-coupling term in $\hat H$ entails
$[ {\hat N}_{\uparrow} ,{\hat H}] \ne 0$ and $ [ {\hat N}_{\downarrow} ,{\hat H}]\ne 0 $. 
Following the variational Hartree-Fock 
approach applied to the two-dimensional gas \cite{sala-njp2017},
Hamiltonian (\ref{H2}) can be reformulated in the mean-field form
where the term ${\hat n}_{\uparrow} {\hat n}_{\downarrow}$ is
expressed as
\be 
{\hat n}_{\uparrow}^+  {\hat n}_{\downarrow}
\simeq 
%
n_{\downarrow} \ {\hat n}_{\uparrow}^+  + n_{\uparrow} \ {\hat n}_{\downarrow}^+ 
- n_{\uparrow} \, n_{\downarrow},
\ee
by using the well-known mean-field (MF) formula for operator 
products \cite{MF1,MF2,MF3}.
This shows how the two spin components are coupled through 
the variational parameters $n_\sigma = \langle {\hat n}_\sigma\rangle$
while the resulting MF Hamiltonian assumes a quadratic form in terms of 
fields ${\hat \psi}_{\sigma}^+$ 
and ${\hat \psi}_{\sigma}$
\be 
{\hat H} =  \int d^3{\bf r} \left\{ \left( {\hat \psi}_{\uparrow}^+ ,  
{\hat \psi}_{\downarrow}^+\right) 
\ {\cal H} \ 
\left( \begin{array}{c} 
{\hat \psi}_{\uparrow} \\
{\hat \psi}_{\downarrow}
\end{array} \right) \right\} - {gn^2\over 4} (1-\zeta^2) L^3 \; ,
\label{HMF} 
\ee
where $L^3$ is the volume of the 3D system and 
\be 
{\cal H} =  
 -\frac{\hbar^2}{2m}\nabla^2  + \frac{gn}{2} {\mathbb I} 
- \frac{gn}{2} \zeta \sigma_z + \frac{\hbar \Omega}{2} \sigma_x\; ,
\label{hsp}
\ee
has been expressed by using Pauli matrices  $\sigma_z$ and $\sigma_x$ with
$\zeta =(n_{\uparrow }-n_{\downarrow})/ n$.
To reduce such Hamiltonian to the diagonal form one first express ${\hat \psi}_{\sigma}^+$ 
and ${\hat \psi}_{\sigma}$ through the standard representation involving momentum-mode operators
${\hat b}^+_{{\bf k} \sigma}$ and ${\hat b}_{{\bf k} \sigma}$ and then implement the unitary transformations
\be
{\hat b}_{{\bf k} \up} = C \ {\hat b}_{{\bf k}, +} + S \ {\hat b}_{{\bf k}, -}
, \quad{\hat b}_{{\bf k} \dn} = C \ {\hat b}_{{\bf k}, -} - S \ {\hat b}_{{\bf k}, +},
\label{tran}
\ee
where the new ladder operators ${\hat b}_{{\bf k},s}$ and ${\hat b}_{{\bf k},s}^{+}$ have been defined and 
$C= \cos \phi/2$ and $S= \sin \phi/2$. The diagonal form is achieved when the angle $\phi$ is given by
\be
{\rm tg} \phi = {\hbar \Omega}/(\zeta {gn})\ .
\label{tg}
\ee 
As a consequence, Hamiltonian (\ref{H2}) can be written as 
\be 
{\hat H} = - {gn^2\over 4} (1-\zeta^2) L^3 + 
\sum_{\bf k} \sum_{s=-1,1} E_{{\bf k},s} 
\ {\hat b}_{{\bf k},s}^{+} \ {\hat b}_{{\bf k},s} \; , 
\label{h-mf}
\ee 
in which 
the energy eigenvalues read
\be 
E_{{\bf k},s} =  \frac{ \hbar^2k^2}{2m} + \alpha_s 
\label{energia}
\ee
with 
\be 
\alpha_s =  \frac{g}{2} {n} + {s\over 2} \; R \; , \quad\quad\quad 
R = \sqrt{ {g^2n^2} \zeta^2  + {\hbar^2 \Omega^2} } . 
\label{energia-pezzetto}
\ee
A couple of parameters emerge from this approach 
\be 
n=n_{\uparrow} +n_{\downarrow}, \qquad
\zeta ={n_{\uparrow }-n_{\downarrow} \over n} ,
\ee
which represent the average total number density 
and the population imbalance, respectively. 
According with the current Hartree-Fock MF scheme, 
$\zeta$ represents the variational parameter of the model whose value can 
be determined by minimizing the total energy. 
At fixed total density ${n}$, the natural range of the 
imbalance parameter turns out to be $\zeta\in [-1,1]$. 


\section{Ground-state energy and phase diagram}
\label{sez3}
In order to obtain an explicit analytic formula for both the total energy 
and the total particle number it advantageous to recast these quantities 
into the form 
\be
E =  {gn^2\over 4} (\zeta^2-1) L^3 + \sum_s \frac{L^3}{8\pi^3 }\int \! d^3{ k}\; E_{{\bf k},s} n_{s}(k) \; , 
\ee
%
\be
N = 
\sum_s \frac{L^3}{(2\pi)^3 }\int \! d^3 k  \; n_{s} (k) \, , 
\ee
respectively, in which the continuum limit has been performed on the summations involving 
the momentum vector ${\bf k}$.  The use of the  zero-temperature fermionic densities
$n_{s}(k) = \langle {\hat b}_{{\bf k},s}^{+} \ {\hat b}_{{\bf k},s} \rangle $ 
$= \theta \bigl(\mu - E(k,s) \bigr) = 
\theta \bigl(\mu -(q + \alpha_s ) \bigr) $
with $\theta(x)$ the Heaviside step function and $q= \hbar^2 k^2/(2m)$, 
provides, after straightforward calculations,  the total-energy density 
\bea
{\cal E} = \frac{gn^2}{4} (\zeta^2- 1) 
&+& \!\frac{2}{(2\pi)^2}\left (\frac{2m}{\hbar^2}\right )^{3/2} \times     
\nonumber
\\
\sum_{s=\pm}  \Bigl [ \frac{(\mu -\alpha_s)^{5/2}}{5} 
&+& \! \frac{\alpha_s}{3} (\mu -\alpha_s)^{3/2} \Bigr ]\ \theta(\mu -\alpha_s) ,  
\label{ener} 
\eea
where ${\cal E}=E/L^3=\langle {\hat H}\rangle/L^3$,
and the total number density
\be
n = \frac{1}{6\pi^2} \left (\frac{2m}{\hbar^2}\right )^{3/2} 
\Bigl [ (\mu -\alpha_+)^{3/2} \theta(\mu -\alpha_+) 
+
(\mu -\alpha_-)^{3/2} \theta(\mu -\alpha_-) \Bigr ] ,
\label{den}
\ee
where $n=N/L^3=\langle {\hat N}\rangle/L^3$
and $\alpha_{-}$ and $\alpha_{+}$ are given by Eq. (\ref{energia-pezzetto}).
In Eqs. (\ref{ener}) and (\ref{den}) both ${\cal E}$ and $n$ depend 
on the chemical potential $\mu$ and the population imbalance $\zeta$. 
Also, we note that the definition of Fermi momentum 
$k_F = (3 \pi^2 n)^{1/3}$ for a two-component gas \cite{pilati} 
is naturally involved in (\ref{den})
which takes the form
\be
\left (\frac{\hbar^2 k^2_F}{2m} \right )^{3/2} = \frac{1}{2}
\left  [ (\mu -\alpha_+)^{3/2} \theta(\mu -\alpha_+) 
+
(\mu -\alpha_-)^{3/2} \theta(\mu -\alpha_-) \right ].
\label{Kden}
\ee
%
\begin{figure}[ht]
\begin{center}
\includegraphics[clip,width= 0.4
\columnwidth 
]{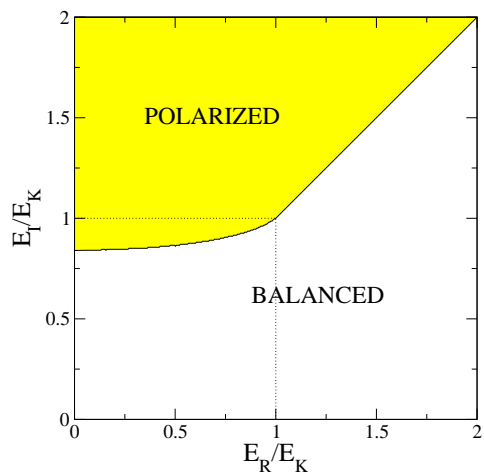}
\caption{Phase diagram of in the plane $(E_R/E_K,E_I/E_K)$, 
where $E_K$ is the kinetic energy per particle, 
$E_R=\hbar \Omega$ is the Rabi energy, and 
$E_I=gn$ is the interaction energy per particle. 
Balanced means $\zeta=0$ and polarized means $\zeta\neq 0$.}
\label{fig1}
\end{center}
\end{figure} 
%
In our approach (see also \cite{sala-njp2017}), 
we first write ${\cal E}$ as function of $n$ 
and $\zeta$ and then we find the value of $\zeta$ which minimizes the 
energy density ${\cal E}$ at fixed total number density $n$. 
The procedure is repeated for different values of interaction strength $g$ 
and Rabi frequency $\Omega$. 
As suggested by equations (\ref{hsp}), (\ref{ener}) and (\ref{Kden}), 
we define the characteristic energy scales
\be
E_R=\hbar \Omega
\quad E_I=gn , \quad 
E_K= \frac{\hbar^2}{2m} (6\pi^2n)^{2/3} = 2^{2/3} E_F ,
\label{defin}
\ee
representing the Rabi energy, the interaction energy per particle, and the kinetic 
energy per particle, respectively. To favor the comparison with the Fermi-gas literature
the latter has been also written in terms of
the Fermi energy $E_F={\hbar^2 k^2_F}/{(2m)}$. The extra factor $2^{2/3}$
bears memory of the fact that we deal with a two-component gas.

The main result of the paper is shown in Fig. \ref{fig1}, 
where we report the balanced-polarized phase diagram of the system in the plane $(E_R/E_K,E_I/E_K)$. 
The solid line is the critical curve of Stoner instability \cite{stoner}, 
where the uniform balanced configuration becomes unstable. 
%
%
%
The derivation of the phase diagram of Fig. \ref{fig1} 
is discussed in detail for the two regimes $ \alpha_- < \mu <\alpha_+$ and $ \alpha_\pm < \mu $.

\subsection{First regime} 
For $\alpha_- < \mu <\alpha_+$ formula (\ref{Kden}) 
reduces to
%
\be
\mu - \alpha_- =  E_K
\label{mu1}
\ee
while the energy density becomes
\be
{\cal E} = \frac{n}{4} E_I (\zeta^2-1) 
+ 3n \left [  \frac{E_K}{5}
+ 
\frac{E_I}{6} - \frac{1}{6} \sqrt{ E_I^2 \zeta^2 +E_R^2} 
\right ].
\ee
The characteristic energies  (\ref{defin}) have been used. 
Then, one easily finds that the extremal values of $\cal E$ are determined 
by the condition
\be
0 = \frac{n}{2} E_I  \zeta  \left [ 1 
- \frac{E_I}{ \sqrt{ E_I^2 \zeta^2 + E_R^2} }  
\right ] .
\label{equaz0}
\ee
The latter exhibits the two solutions
$1) \,\, \zeta = 0$ (for $E_I < E_R$) and
$2) \,\, \zeta = \sqrt{1 -{E_R^2}/{E_I^2}}$ (for $E_I > E_R$),
with the energies
\be
{\cal E}_1 = n \;  \left ( \frac{E_I}{4}  +   \frac{3 E_K}{5} - 
\frac{E_R}{2}  \right ), \qquad
{\cal E}_2 =  -\frac{E_R^2}{4g} +   3n  \frac{E_K}{5} ,
\ee
associated to $\mu_1 = E_K + {(E_I- E_R)}/{2}$, and
$\mu_2 = E_K $, respectively.
One easily checks that ${\cal E}_2 = {\cal E}_1$  and $\mu_1 = \mu_2$  
at the critical value $E_I =E_R$.
The inclusion of the constraint  $\alpha_- < \mu < \alpha_+$ (characterizing the current case)
rewritten as $E_K =\mu -\alpha_-< \alpha_+ -\alpha_-$ (see (\ref{mu1})) entails, in turn,
\be
E_K < \alpha_+ -\alpha_- = 
\sqrt{ E_I^2 \zeta^2 +E_R^2}  .
\ee
This new constraint allows one to define the exact range of validity of the two solutions of
equation (\ref{equaz0}), giving
\bea
&1)& \, \zeta = 0 \qquad {\rm for}\,\,\, E_I < E_R, \,\,\,
E_K  < E_R ,
\label{sol1}
\\
&2)& \, \zeta = \sqrt{1 -\frac{E_R^2}{E_I^2}}
\quad {\rm for}\, \, \, E_R < E_I,
\,\,\,  E_K <  E_I .
\label{sol2}
\eea
These well reproduce, in Fig. \ref{fig1}, the balanced-phase domain and polarized-phase 
domain, respectively, placed outside the squared box $E_I, E_R < E_K$.

\subsection{Second regime}
This case is characterized by $\alpha_\pm <  \mu$. Then, formula (\ref{den})  becomes
\begin{equation}  
E^{3/2}_K = (\mu -\alpha_+)^{3/2}   + (\mu -\alpha_-)^{3/2} ,
\label{n1} 
\end{equation}
where $E_K$ has been defined in equation (\ref{defin}). 
The constraint (\ref{n1}),  in which $\alpha_\pm = (gn\pm R)/2$, causes 
the implicit dependence of $\mu$ from $R$, and thus
from the mean-field parameter $\zeta$ contained in $R$. By introducing the new variables
\be
\xi = (\mu -E_I/2) - R/2, \qquad \eta = (\mu -E_I/2) + R/2,
\label{xieta}
\ee
the constraint (\ref{n1}) becomes
\begin{equation}
E_K^{3/2} = \xi^{3/2}  + \eta^{3/2},
\label{const}
\end{equation}
where the fact that $\mu$ and $R$ (depending on $\zeta$) are independent 
variables implies that $\xi$ and $\eta$ are, in turn, independent variables.
Equation (\ref{const}) entails that the natural range of $\xi$ and $\eta$ is $\xi, \eta \in [0, E_K]$.
In parallel, the use of variables 
$\xi$ and $\eta$ and of  the explicit expressions 
of $\alpha_\pm$ in terms of $R$ and $ng =E_I$ in the energy density (\ref{ener}) (in which, now,  $\theta (\mu -\alpha_\pm ) = 1$) gives
\be
{\cal E} = \frac{n}{4} E_I (\zeta^2+1) 
+ \frac{3n}{E_K^{3/2} } 
\left [  \frac{\xi^{5/2} +\eta^{5/2} }{5}  + \frac{R}{6} (\xi^{3/2}- \eta^{3/2})  \right ].
\label{ener2}
\ee 
Then, by exploiting constraint (\ref{const}), 
written in the form
$\eta = (E_K^{3/2} -\xi^{3/2})^{2/3}$ in order to eliminate $\eta$ from $\cal E$, 
the energy density becomes
\begin{equation}
{\cal E} = \frac{n}{4} E_I (\zeta^2+ 1) + \frac{3n}{E_K^{3/2} } \; G(\zeta) ,
\label{enerG}
\end{equation}
where
\be
G = G[R, \xi(R)]
=\frac{1}{5} \xi^{5/2}  + \frac{1}{5}  \left ( E_K^{3/2}-\xi^{3/2} 
\right )^{5/3}  
+ \frac{R}{6}  \left ( 2\xi^{3/2} - E_K^{3/2} \right )  .
\label{eqG}
\ee
Its derivative ${dG}/{d \zeta}$ can be easily calculated and, thanks to the identity  $R = ( E_K^{3/2}-\xi^{3/2} )^{2/3} -\xi $,
its final form is found to be
\begin{equation}
\frac{dG}{d \zeta} =  
\frac{ 2\xi^{3/2} -E_K^{3/2} }{6}   \frac{d R}{d \zeta} =  
\left ( 2\xi^{3/2} -E_K^{3/2} \right )    \frac{E_I^2 \zeta }{6 R} 
\label{derfin}
\end{equation}
where $R = \sqrt{ E_I^2 \zeta^2 + E_R^2}$ has been used. 

In conclusion, in the current regime where $\alpha_\pm <  \mu$, one finds 
the energy-density derivative
\begin{equation}
\frac{d \cal E}{d \zeta} =
\frac{n E_I}{2R} \zeta 
\left( R  + \frac{2E_I}{E_K^{3/2} }  \xi^{3/2}  -E_I \right ) = 0
\label{enerder}
\end{equation}
providing the explicit solution
\begin{equation}
\zeta =0 , 
\label{1solution}
\end{equation}
and the implicitly-defined solution
\begin{equation}
R  + \frac{2E_I}{E_K^{3/2} }  \xi^{3/2}   -E_I =0 .
\label{2solution}
\end{equation}
Equation (\ref{2solution}) must be solved numerically. It is worth noting, however, that
in the limit $\xi/E_K \ll 1$ (one should remind that this inequality 
corresponds to $\mu$ larger than,  but very close to $\alpha_+$) 
the term $(\xi/E)^{3/2}$ can be neglected and $R -E_I \simeq 0$ entails 
$
\zeta \simeq \pm \sqrt{1 -E_R^2/E_I^2} 
$ 
with $E_R < E_I$. By keeping the approximation  $(\xi/E)^{3/2} $ $\simeq 0$, 
the second derivative gives
\be
\frac{d^2 \cal E}{d \zeta^2} 
\simeq
\frac{nE_I}{2R} \left ( R  -E_I \right ) + \frac{nE_I^2 }{2R^3} \zeta^2  
\ee
showing that the first solution one finds, $\zeta=0$, is a minimum for $R=E_R > E_I$ 
and a maximum for $R=E_R < E_I$,
whereas the second solution $\zeta \simeq $ $\pm 
\sqrt{1 -E_R^2/E_I^2 } $
is always a minimum provided that $E_R < E_I$. 
In the box $E_I/E_K \le 1$, $E_R/E_K \le 1$ (the region in Fig. \ref{fig1} corresponding
to the current regime $\alpha_\pm <  \mu$) these results are in agreement with the presence 
of both the polarized phase ($E_I >E_R$ with  $E_I/E_K >1$) and the balanced 
phases ($E_I <E_R$ with  $E_R/E_K >1$) highlighted in regime 
$\alpha_- < \mu < \alpha_+$, and essentially represent the prolongation 
of such phases inside this box. 
%
\begin{figure}[ht]
\begin{center}
\includegraphics[clip,width= 0.5
\columnwidth
]{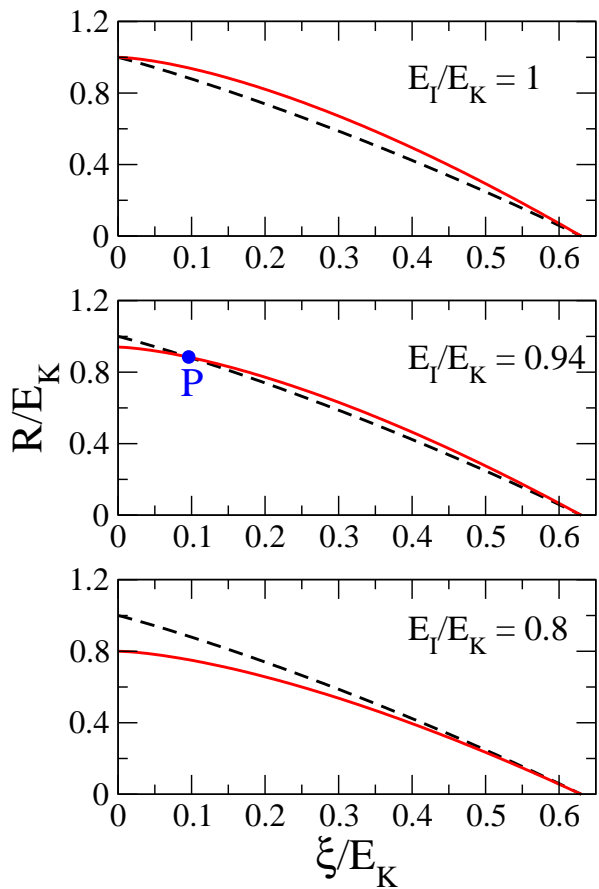}
\caption{(Color Online) 
Curves describing $R/E_K$ vs $\xi/E_K$
according to Eqs. (\ref{R1}) and (\ref{R2}), 
which are, respectively, the solid line and the dashed line. 
The three panels correspond to different values of the adimensional 
interaction strength $E_I/E_K$. Only in the middle panel 
there is an intersection point $P$.}
\label{fig2}
\end{center}
\end{figure}
%

\section{Phase boundary for $E_R < E_K$, $E_I < E_K$}
\label{sez4}

The phase boundary inside the 
box $x= E_R/E_K < 1$, $y =E_I/E_K < 1$ of Fig. \ref{fig1} is found by numerically 
calculating the values of $\zeta$ determined by means of equation (\ref{2solution})
to minimize energy (\ref{enerG}). 
Unlike the 2D case (see \cite{sala-njp2017}), the separatrix between the 
polarized phase and the balanced phase is not a simple straight line 
intersecting the origin $x=y=0$ and bisecting the portion 
of plane $y \ge 0$, $x \ge 0$.
This is due to the form of the equation (\ref{2solution}) rewritten as
\begin{equation}
R  = E_I \left ( 1- {2\xi^{3/2}}/{E_K^{3/2} }   \right ), 
\label{R1}
\end{equation}
whose solution requires the use of the auxiliary equation
\begin{equation}
R = -\xi + \left ( E_K^{3/2}-\xi^{3/2} \right )^{2/3}.
\label{R2}
\end{equation}
The latter, exhibiting a manifest nonlinear character, embodies the constraint 
(\ref{const}), which due to $\eta - \xi = R$ (see equations (\ref{xieta}))
implicitly defines $\xi$ as a function of $R$.

Despite the absence of an analytic explicit solution, 
some useful information can be extracted from such equations by representing 
them on the region of plane $(R, \xi)$ with 
$R, \xi \ge 0$ (see Fig. \ref{fig2}). 
With 1 and 2 referred to the functions (\ref{R1}) and (\ref{R2}), respectively,
one observes that $R_1(\xi) = R_2(\xi) =0$ for $\xi= E_K /2^{2/3}$, 
while $R_1 (0) = E_I$ and $R_2(0) = E_K$. 
In addition to $R_1(\xi) = R_2(\xi) =0$, equations (\ref{R1}) and (\ref{R2}) feature a
second solution.
This follows from the fact that $R_1(\xi)$ has zero derivative 
at $\xi =0$ whereas $R_2(\xi)$ is strongly decreasing for $\xi \simeq 0$. 
Then, for $E_I$ close enough to $E_K$, the two curves must intersect to 
each other in a single point. The derivatives $R_1'(\xi)$ and $R_2'(\xi)$
%
at $\xi_0 = E_K /2^{2/3}$
provide the critical condition

$$
R_1'(\xi_0) = - \frac{3 E_I}{ E_K 2^{1/3} } 
\le R_2'(\xi) 
= -2,
$$
stating that for $E_I \ge (2^{4/3}/3) E_K$ the two curves 
exhibit an intersection point P in the plane $(R, \xi)$, 
shown in the middle panel of Fig. \ref{fig2}.
In parallel, the top panel in  Fig. \ref{fig2} shows how,
for $E_I /E_K\equiv 1$, P reaches the point ($R= E_K$, $\xi=0$), 
while, for $E_I /E_K\equiv 2^{4/3}/3 \simeq 0.8399$, the intersection point P tends to 
the point described by ($R=0$, $\xi_0 = E_K /2^{2/3}$). 
Finally, the bottom panel in  Fig. \ref{fig2}
highlights how, for $E_I /E_K < 2^{4/3}/3$, no intersection 
survives.

By denoting with $A$ the portion of the yellow phase in the box 
$E_I/E_K \le 1$, $E_R/E_K \le 1$ of Fig. \ref{fig1},
one discovers that the interval $I = \{ E_I/E_K\in [2^{4/3}/3, 1] \}$ on the
vertical axis actually represents the parameter range in which the system is in the polarized 
phase $|\zeta| \ne 0$ within the box. 
Particularly, any value $R_* = R(E_I, E_K)$ representing a solution of 
equations (\ref{1solution}) and (\ref{2solution}) for $E_I/E_K \in I$
allows one to determine $|\zeta|$ through the formula 
$R = ( E_I^2 |\zeta|^2 +E_R^2 )^{1/2}$. 
The phase boundary confining $A$ from below is found as follows. 
Consider a specific horizontal 
(straight) line $y= E_I/E_K = const $ (with $E_I$, $E_K$ such that 
$E_I/E_K \in I$), crossing $A$ and intersecting the $A$ boundary.
The resulting $|\zeta|$ is given by  $|\zeta| = (R_*^2- E_R^2)^{1/2} /E_I $. 
The two solutions $|\zeta | =R_*/E_I $, $E_R = 0$ and  $\zeta =0 $, $E_R = R_*$ 
can be shown to be associated to the two intersection points of 
$y= E_I/E_K $ with $I$ (on the vertical axis) and the $A$ boundary, 
respectively. Intermediate values of $|\zeta|$ describe the points of the 
straight line inside $A$.

We conclude by noting how the Stoner condition $k_F = \pi/(2a)$, 
where $a$ is the scattering length of the interaction parameter 
$g = 4\pi^2 \hbar^2 a/m$, for the transition to the polarized phase 
is recovered for $E_R =0$. In Fig. \ref{fig1}, the critical point 
corresponds to the lowest value $E_I/E_K = 2^{4/3}/3$ of the interval 
$I$. Rewriting $E_I= gn$ as $E_I = g k_F^3/(3\pi^2)$ 
the Stoner result is easily found. 

We stress that the phase diagram of Fig. 1 is meaningful also 
in the presence of an external trapping potential $U({\bf r})$, where 
the total number density becomes space dependent, i.e. $n=n({\bf r})$.  
In this case, within the local density approximation, 
the fermionic cloud is fully balanced if $n({\bf r})^{1/3} < 
(2^{4/3}/3) \hbar^2 (6\pi^2)^{2/3}/(8\pi a)$ for any ${\bf r}$, 
while for $\Omega \neq 0$ the fermionic cloud can have an 
unbalanced region ${\cal D}$ under the condition 
$n({\bf r})^{1/3} > (2^{4/3}/3) (6\pi^2)^{2/3}/(8\pi a)$ with ${\bf r}
\in {\cal D}$. Due to the experimental 
flexibility of scattering and Rabi couplings, we believe 
our prediction can be tested by using available 
experimental setups with, for instance, $^6$Li atoms on a 
spherically-symmetric cloud of about 100 micron of radius.


\section{Spin polarizations}
\label{sez5}

The information emerging from our analysis and the phase diagram 
of Fig. \ref{fig1} can be effectively represented by considering 
suitable indicators of the system polarization state. 
Let us consider the $z$ and $x$ components of the total spin operator
\bea
{\hat S}_z &=& 
\frac{1}{2} \int d^3{\bf r} \ {\hat \Psi}^+ \sigma_z {\hat \Psi} =
\frac{1}{2} \int d^3{\bf r} \ \Bigl( {\hat \psi}^+_{\up} {\hat \psi}_{\up}  
- {\hat \psi}^+_{\dn} {\hat \psi}_{\dn} \Bigr) 
\nonumber 
\\
&=& 
\frac{1}{2} \sum_{\bf k} \Bigl( {\hat n}_{{\bf k} \up} - 
{\hat n}_{{\bf k} \dn} \Bigr),
\eea
\bea 
{\hat S}_x &=& \frac{1}{2} \int d^3{\bf r} 
\ {\hat \Psi}^+ \sigma_x {\hat \Psi} =
\frac{1}{2} \int d^3{\bf r} \ \Bigl( 
{\hat \psi}^+_{\up} {\hat \psi}_{\dn} + {\hat \psi}^+_{\dn} 
{\hat \psi}_{\up} \Bigr)
\nonumber 
\\
&=&
\frac{1}{2} \sum_{\bf k} \Bigl(  {\hat b}^+_{{\bf k} \up} 
{\hat b}_{{\bf k} \dn} 
+ {\hat b}^+_{{\bf k} \dn} {\hat b}_{{\bf k} \up} \Bigr),
\eea 
respectively, in which
%
${\hat \Psi}({\bf r})= ( {\hat \psi}_{\up}({\bf r}) , {\hat \psi}_{\dn}({\bf r}) )$ is the 
two-component spinor. 
By using transformations (\ref{tran}) one finds
%
\be
{\hat S}_z = \sum_{\bf k} \Bigl [ 
\frac{\cos \phi}{2} \Delta n_{\bf k}
+
\frac{\sin \phi}{2}  ({\hat b}^+_{{\bf k}, +} {\hat b}_{{\bf k}, -} 
+ 
{\hat b}^+_{{\bf k}, -} {\hat b}_{{\bf k}, +} )
\Bigr ] ,
\label{osz}
\ee
\be
S_x 
= 
\sum_{\bf k} \Bigl [
\frac{\cos \phi}{2}  ({\hat b}^+_{{\bf k}, +} {\hat b}_{{\bf k}, -} 
+ {\hat b}^+_{{\bf k}, -} {\hat b}_{{\bf k}, +} ) 
-\frac{\sin \phi}{2} \Delta n_{\bf k}
\Bigr ] ,
\label{osx}
\ee
%
where the identities $\cos \phi \equiv \zeta  E_I/R$, $\sin \phi \equiv E_R/R$ 
follow from equation (\ref{tg}), and
\be
\Delta n_{\bf k} \equiv {\hat n}_{{\bf k}, +} - {\hat n}_{{\bf k}, -}, \quad R = \sqrt {g^2 n^2 \zeta^2 +\hbar^2 \Omega^2}.
\ee
%
If state
$| E_0\rangle = $ $
\prod_{\bf k} | n_{{\bf k}, +} \rangle | n_{{\bf k}, -} \rangle$
is the ground state satisfying $H_{mf} | E_0\rangle = E_0 | E_0\rangle$, then the expectation values
\be
\langle E_0 | {\hat S}_z | E_0\rangle =  
\frac{\zeta E_I}{2 R} \sum_{\bf k} \langle E_0 | \Delta n_{\bf k} | E_0\rangle,
\label{sz}
\ee
\be
\langle E_0 | {\hat S}_x | E_0\rangle = \frac{E_R}{2R}  \sum_{\bf k} \langle E_0 | \Delta n_{\bf k}  | E_0\rangle,
\label{sx}
\ee
can be derived  from (\ref{osz}) and (\ref{osx}). Eq. (\ref{sz} )  can be rewritten 
in the more expressive way
\be
\langle E_0 | {\hat S}_z | E_0\rangle = 
\sum_{\bf k} \langle E_0 
| \frac{{\hat n}_{{\bf k} \up} - {\hat n}_{{\bf k} \dn}}{2} | E_0\rangle 
= \frac{L^3}{2} (n_\up -n_\dn)
\label{nn}
\ee
while the summation in the right-hand side of identities (\ref{sz}) and (\ref{sx}) can be shown 
to assume the form
\bea
\sum_{\bf k} \langle E_0 | \Delta n_{\bf k} | E_0\rangle &=& 
\frac{n L^3}{E_K^{3/2}} \Bigl [ 
(\mu -\alpha_+)^{3/2} \theta (\mu -\alpha_+)  
\nonumber 
\\ 
&-& (\mu -\alpha_-)^{3/2} \theta (\mu -\alpha_-) \Bigr ].
\label{pan} 
\eea
%
Introducing equations (\ref{nn}) and (\ref{pan}) in (\ref{sz}) 
enables us to obtain the self-consistent equation characterizing our MF approach 
\be
\zeta \left [ 1 + \frac{E_I}{R E_{K}^{3/2}} \sum_{ p =\pm} \; p \left  (\mu -\alpha_p \right )^{3/2} 
\theta \left (\mu -\alpha_p \right )  \right] = 0 .
\label{sce}
\ee
Thanks to the latter one has

$$
\frac{R E_K^{3/2}}{ng} = (\mu -\alpha_-)^{3/2} \theta (\mu -\alpha_-)
-  (\mu -\alpha_+)^{3/2} \theta (\mu -\alpha_+) . 
$$
Therefore, equation (\ref{pan}) takes the form
$\sum_{\bf k} \langle E_0 | \Delta n_{{\bf k}} | E_0\rangle $ $= - {n L^3 R}/{(gn)} $
and the average values $\langle {\hat S}_z\rangle $ and 
$\langle {\hat S}_x\rangle $ are simply given by
\be
\langle {\hat S}_z\rangle =  {\zeta N}/{2} 
, \quad 
\langle {\hat S}_x \rangle = - {N E_R}/{(2 E_I)} \; . 
\label{expv}
\ee
%
The first formula $\langle {\hat S}_z\rangle = N \zeta/2$ 
confirms the validity of the Hartree-Fock MF scheme showing how our variational Hartree-Fock method 
reproduces the correct description of the average polarization per particle of component ${\hat S}_z$
through the variational parameter $\zeta$. In the yellow area of Fig. (\ref{fig1}), $\langle {\hat S}_z\rangle$
ranges from $N/2$ when $x=E_R/E_K=0$ to  $\langle {\hat S}_z\rangle=0$ at any point of phase separatrix.
%
%
On the other hand, the second formula in equation (\ref{expv}), describing the 
spin component ${\hat S}_x$
suggests that $\langle {\hat S}_x \rangle \in [0, -N/2]$ for ${E_R} < E_I$. 
In particular, one has $\langle {\hat S}_x \rangle = 0$ 
for $E_R=0$, and $\langle {\hat S}_x \rangle = -N/2$  for $E_R =E_I$. 
The latter is maintained for $E_I < {E_R}$, consistent with the fact that, in the
extreme case when $E_I =0$, only the Rabi coupling survives in $H$.
This confirms that the average polarization per particle of component ${\hat S}_x$
crucially depends on the interplay between the Rabi energy per particle $E_R=\hbar \Omega$ 
and the interaction energy per particle $E_I=gn$. 
%
\section{Conclusions}

We have investigated a 3D repulsive Fermi gases formed by two pseudo-spin 
components corresponding to the two Rabi-coupled hyperfine states.
We have shown that Stoner instability and polarization of component ${\hat S}_z$
are induced by the interplay between Rabi coupling and repulsive interaction. 
Our results, obtained by adopting a variational Hartree-Fock mean-field scheme, 
are similiar to the ones of the 2D Fermi gas \cite{sala-njp2017}. 
However, contrary to the 2D case, in three spatial dimensions both the ground-state
energy and the total fermion number of the uniform system feature a very complex nonlinear
dependence on the chemical potential and the population imbalance. 
By performing analytical calculations we have successfully obtained the ground-state
energy as a function of the total number density for the different regimes of the 
chemical potential. We have then derived the zero-temperature balanced-to-polarized 
phase diagram of the uniform system. 
This has enabled us in describing the competition of the Rabi coupling 
with the interaction $g$ in determining the pseudo-spin polarization of the fermionic 
gas and, more specifically, when the polarization of one of the two components 
${\hat S}_z$ and ${\hat S}_x$ prevails. 
In particular, we have obtained that a Rabi coupling strong enough implies 
that $|\langle {\hat S}_x \rangle|$ reachs its maximum value while 
$\langle {\hat S}_z \rangle$ becomes negligible.
In the opposite case, the critical value $E_I = 2^{4/3} E_K/3$, characterizing
the transition to 
a spin-polarization dominated by ${\hat S}_z$ when $E_R=0$ (in this case 
$\langle {\hat S}_z \rangle$ reachs its maximum value) 
has been shown to reproduce the well-known criterion for the Stoner instability. 
Beyond-mean-field quantum fluctuations can reduce the critical interaction 
strength of Stoner instability \cite{pilati}-\cite{conduit1} but we expect that 
this effect is quite small in three dimensions while it could be larger in two 
spatial dimensions. Therefore we believe that our 3D theoretical predictions 
could be a reliable and useful guide for future experiments.

\medskip

\noindent
%


\medskip

\noindent
{\bf Acknowledgments}
LS acknowledges for partial support the 2016 BIRD project 
"Superfluid properties of Fermi gases in optical potentials" of the 
University of Padova. 

%
%


\section*{References}

\begin{thebibliography}{99}

\bibitem{so-bose} Lin Y J, Jimenez-Garcia K and Spielman I B 2011 
{Nature} {\bf 471} {83}

\bibitem{so-bose2} 
Zhang J-Y, Ji S-C, Chen Z, Zhang L, Du Z-D, Yan B, 
Pan G-S, Zhao B, Deng Y-J, Zhai H,  Chen S and Pan J W 2012  
{Phys. Rev. Lett.} {\bf 109} {115301}

\bibitem{so-fermi1}
Wang P, Yu Z-Q, Fu Z, Miao J, Huang L, Chai S, Zhai H and Zhang J 
2012 {Phys. Rev. Lett.} {\bf 109} {095301}

\bibitem{so-fermi2}
Cheuk L W, Sommer A T, Hadzibabic Z, Yefsah T, Bakr W S and Zwierlein M W 
2012 {Phys. Rev. Lett.} {\bf 109} {095302}

\bibitem{stringa1} Li Y, Pitaevskii L P and Stringari S 2012 
{Phys. Rev. Lett.} {\bf 108} {225301}

\bibitem{stringa2}
Martone G I, Li Y, Pitaevskii L P and Stringari S 2012
{Phys. Rev. A} {\bf 86} {063621}

\bibitem{burrello} Burrello M and Trombettoni A 2011 
{Phys. Rev. A} {\bf 84} {043625}

\bibitem{so-solitons1} Merkl M, Jacob A, Zimmer F E, Ohberg P and Santos L 
2010 {Phys. Rev. Lett.} {\bf 104} {073603}

\bibitem{so-solitons2} Fialko O, Brand J and Z\"ulicke U 2012 
{Phys. Rev. A} {\bf 85} {051605}

\bibitem{so-solitons3} Liao R, Huang Z-G, Lin X-M and Liu W-M 2013 
{Phys. Rev. A} {\bf 87} {043605}

\bibitem{so-bright1} Achilleos V, Frantzeskakis D J, Kevrekidis P G 
and Pelinovsky D E 2013 Phys. Rev. Lett. {\bf 110}, 264101 

\bibitem{so-bright2} Xu Y, Zhang Y, and Wu B 2013 Phys. Rev. A \textbf{87} 013614 

\bibitem{so-bright3} Salasnich L and Malomed B A 2013 Phys. Rev. A \textbf{87} 063625 

\bibitem{shenoy1} Vyasanakere J P and Shenoy V B 2011 Phys. Rev. B {\bf 83} 094515 

\bibitem{shenoy2} Vyasanakere J P, Zhang S and Shenoy V B 2011 
Phys. Rev. B {\bf 84} 014512  

\bibitem{gong} Gong M, Tewari S and Zhang C 2011 Phys.
Rev. Lett. {\bf 107} 195303  

\bibitem{hu} Hu H, Jiang L, Liu X-J and Pu H 2011 Phys. Rev.
Lett. {\bf 107} 195304 
 
\bibitem{iskin} Iskin M and Subasi A L 2011 Phys. Rev. Lett. {\bf 107} 050402 

\bibitem{dms1} Dell'Anna L., Mazzarella G., and Salasnich L., Phys. 
Rev. A 2011 {\bf 84} 033633 

\bibitem{dms2} Dell'Anna L., Mazzarella G., and Salasnich L., 
Phys. Rev. A 2012 {\bf 86} 053632 

\bibitem{sa} Han L. and Sa de Melo C. A. R., Phys. Rev. A 2012 {\bf 85} 
011606(R)  

\bibitem{chen} Chen G., Gong M. and C. Zhang C., Phys. Rev. A 2012 {\bf 85}
013601 

\bibitem{zhou2} Zhou K. and Zhang Z., Phys. Rev. Lett. 2012 {\bf 108}
025301 

\bibitem{yangwan} Yang X. and S. Wan S.,  Phys. Rev. A 2012 {\bf 85} 023633 

\bibitem{roati} Valtolina G, Scazza F, Amico A, Burchianti A, Recati A, 
Enss E, Inguscio M, Zaccanti M and Roati G 2016 {Nature Physics} {\bf 13} {704}

\bibitem{stoner} Stoner E C 1947 {Rep. Prog. Phys.} {\bf 11} {43}

\bibitem{itinero0} Salasnich L, Pozzi B,  Parola A and Reatto L 
2000 {J. Phys. B} {\bf 33} {3943}

\bibitem{itinero1} Jo G-B, Lee Y-R, Choi J-H, Christensen C A, 
Kim T H, Thywissen J H, Pritchard D E and Ketterle W 2009 
{Science} {\bf 325} {1521}

\bibitem{itinero2} Conduit G J,  Green A G and Simons B D 2009 
{Phys. Rev. Lett.} {\bf 103} {207201}

\bibitem{itinero3} Massignan P, Yu Z and Bruun G M 2013 
{Phys. Rev. Lett.} {\bf 110} {230401}

\bibitem{itinero4} 
Ambrosetti A, Lombardi G, Salasnich L, Silvestrelli P L and Toigo F 
2014 {Phys. Rev. A} {\bf 90} {043614}

\bibitem{itinero5} Jiang Y, Kurlov D V, Guan X-W, Schreck F and Shlyapnikov G V 
2016 {Phys. Rev. A} {\bf 94} {2016} 011601(R)

\bibitem{sala-previous} Salasnich L 2013 Phys. Rev. A {\bf 88} 055601 

\bibitem{sala-njp2017} Penna V. and Salasnich L 2017 
{New J. Phys.} {\bf 19} {043018}

\bibitem{MF1} van Oosten D., van der Straten P. and Stoof H. T. C. 
2001 {Phys. Rev. A} {\bf 63} {053601}

\bibitem{MF2} Buonsante P., Penna V. and Vezzani A. 2005 
{Laser Phys.} {\bf 15} {361}

\bibitem{MF3} Buonsante P, Massel F, Penna V and Vezzani A 2009 
{Phys. Rev. A} {\bf 79} {013623}

\bibitem{pilati}
Pilati S, Bertaina G, Giorgini S and Troyer M 2010 
{Phys. Rev. Lett.} {\bf 105} {030405}

\bibitem{duine} Duine R and MacDonald A 2005 
{Phys. Rev. Lett.} {\bf 95} {230403}

\bibitem{conduit0} Conduit G J 2010 {Phys. Rev. A} {\bf 82} {043604}

\bibitem{conduit1} Conduit G J 2013 {Phys. Rev. B} {\bf 87} {184414}

\end{thebibliography}
\end{document}